# Kinetic Monte Carlo Model of Breakup of Nanowires into Chains of Nanoparticles

Vyacheslav Gorshkov[a] and Vladimir Privman[b],*

[a] National Technical University of Ukraine "Igor Sikorsky Kyiv Polytechnic Institute", Kyiv 03056, Ukraine

[b] Department of Physics, Clarkson University, Potsdam, NY 13699, USA

* privman@clarkson.edu

## ABSTRACT

A kinetic Monte Carlo approach is applied to studying shape instability of nanowires that results in their breaking up into chains of nanoparticles. Our approach can be used to explore dynamical features of the process that correspond to experimental findings, but that cannot be interpreted by continuum mechanisms reminiscent of the description of the Plateau-Rayleigh instability in liquid jets. For example, we observe long-lived dumbbell-type fragments and other typical non-liquid-jet characteristics of the process, as well as confirm the observed lattice-orientation dependence of the breakup process of single-crystal nanowires. We provide snapshots of the process dynamics, and elaborate on the nanowire-end effects, as well as on the morphology of the resulting nanoparticles.

## KEYWORDS

## 1. INTRODUCTION

In synthesis of nanoparticles and nanostructures, stability of the products in rather important in many applications. Therefore, breakup of nanowires at temperatures below melting into small fragments that are typically isomeric (even-proportioned) nanoparticles, has recently attracted substantial interest. Experimental[1-9] and theoretical[10-13] works have been published studying this phenomenon, and it has also been argued that the resulting nanoparticle chains can find their own applications, such as in design of optical waveguides.[1] It is tempting to associate this process with Plateau-Rayleigh instability[14,15] that spontaneously develops in liquid jets. It is well known that, due to random fluctuations and/or end effects, surface waves are formed on a liquid jet that, for some spatial wavelengths, grow and cause breakup into droplets. This phenomenon has been studied in numerous works over the years.[16] As a function of the distance, $x$, along the jet, and assuming for simplicity radially-isotropic perturbations, the radius, $r(x)$, can be expanded into Fourier modes, each of which has the form

$$r(x) = r_0 \pm \delta\sin(2\pi x/\lambda), \quad \delta \ll r_0, \tag{1}$$

where $\lambda$ is the mode's wavelength. This causes non-uniformity of the Laplace pressure along the jet. The resulting transport of liquid can be from the narrow to wide regions for certain wavelengths, which results in instability. For liquids of negligible viscosity, for instance, one can show that the most unstable wavelength is

$$\lambda_0 \approx 9.01 r_0, \tag{2}$$

where $r_0$ is the initial jet radius.

Application of similar concepts for nanowire breakup cannot be done too literally, because the physics of the processes that drive the dynamics might be completely different. There could, however, be certain analogies. First, the nano-object surface can actually melt and become liquid-like above some "premelting" (surface melting) temperature, for which the bulk is still solid.[12] Furthermore, surface diffusion of atoms in nanowires has been conjectured[2] to contribute to effects reminiscent of the liquid instability. Generally, higher Gaussian-curvature surface features correspond to lower binding energies of the surface atoms, which for some surface morphologies can be cause instabilities (amplification of thermal fluctuations due to flow of matter) equivalent to that in liquids. The nanowire will then break into droplet-shaped nanoparticles. Interestingly, a continuum model of the effects of surface diffusion was developed,[10] yielding the result

$$\lambda_0 \approx 8.89 r_0, \qquad (3)$$

which is very close to the result in Equation 2. However, our results reported in this article suggest that "fluid dynamics" concepts should be applied with caution to crystalline and polycrystalline nanostructures, surfaces of which can consist of crystalline faces.

As a useful estimate, we note that, if a cylindrical jet is cut into $\lambda_0$-long pieces, the radii of the ultimately approximately spherical nanoparticles will be $R = (3r_0^2 \lambda_0/4)^{1/3}$, and their centers will be separated in space at distances, $\Lambda \approx \lambda_0$. These lengths, $\lambda_0, r_0, \Lambda, R$, can all only be defined and experimentally probed as approximate measures or averages for nanowires and nanoparticles that have crystalline-face surface (as discussed later), and relations between them are approximate. Experimental work on Cu-nanowires[3] found results consistent with $\Lambda \approx 9r_0$. However, another experiment, on Au-nanowires[1] far below the melting point, reported consistently larger values of $R/r_0$ and $\Lambda/r_0$ than those predicted[10] for isotropic surface energy, and furthermore, indicated significant scatter, between 17% and 32%, in the values of these quantities. Similarly, for Ag-nanowires[4] it was found that $\Lambda \approx 9.6 r_0$ exceeds $9r_0$; the nanoparticle effective radii and distance from each other actually fluctuated 10% and 13%, respectively. All this confirms that various mechanisms for nanowire breakup can depend on the specific experimental conditions, and the analogy with liquid is not generally applicable. It is therefore particularly interesting to focus on those mechanisms of the nanowire breakup and resulting morphology of the produced fragments, which are definitely not characteristic of liquids.

Specifically, solid nanowires need not be exactly cylindrical. Most crystalline or polycrystalline nanowire shaped nanostructures are synthesized[17] or otherwise prepared bound by low free-energy faces corresponding to their crystal structure, and they can only approximate a round cylinder. Furthermore, certain non-isomeric fragments were experimentally[1,6,8,9] observed to remain relatively stable over long time scales, especially for low temperatures. These included dumbbell (two-bulb) and even three-bulb structures. We explore such properties in this work, and we also find that, similar to liquids, there are mechanisms by which nanowire ends can drive the instability as breakup propagating into the structure.

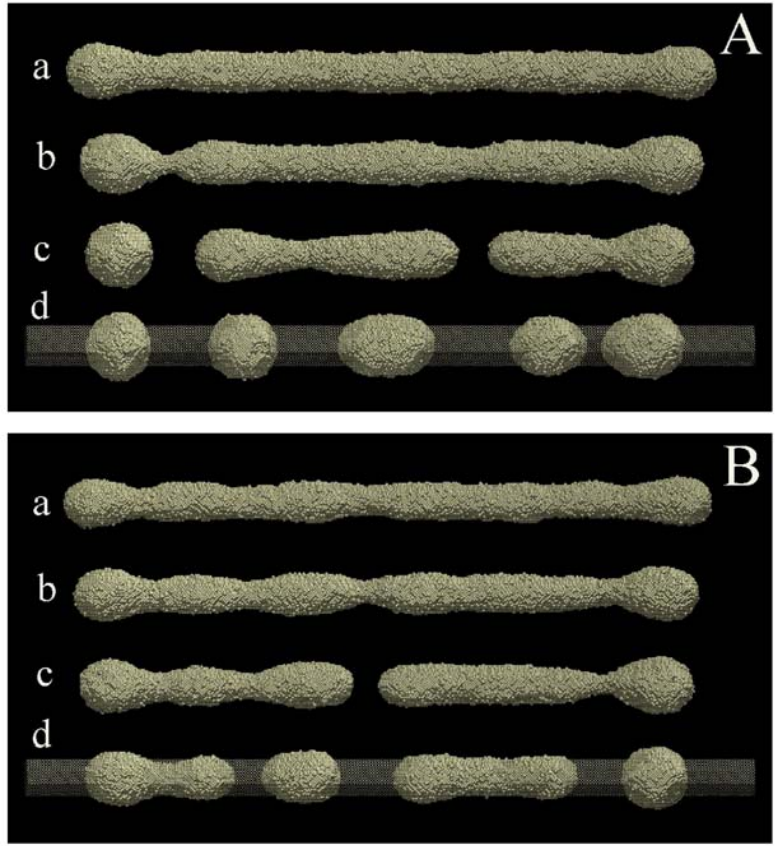

**Figure 1.** Random MC runs showing snapshots at several times, for a nanowire with its axis along the (100)-type orientation of FCC, with model parameters, defined in Sections 2 and 3, $\alpha = 0.9$, $p = 0.725$, $L = 240a$. (A) The effective diameter of the initial nanowire $d = 12a$, with configurations (a), (b), (c), (d) showing the system state after $t = 1.50, 2.40, 3.20, 4.35 \times 10^6$ MC time steps, respectively; (B) $d = 11a$, with configurations labelled (a), (b), (c), (d) showing the system at $t = 1.20, 1.80, 2.10, 2.5 \times 10^6$. The images labelled (d) also outline the initial nanowire, which was bound by four (100)-type FCC faces, and by four (110)-type faces. For later times, the fragments rounded up, but no additional breakup of the shown nanoparticles was observed, including the non-isomeric "dumbbell" ones seen in (B), image (d).

We utilize a kinetic Monte Carlo (MC) model developed and successfully applied[18-24] to mesoscale modeling of processes of synthesis, growth, sintering, and recently also of dissolution of nanoparticles. The prototypical structures were noble-metal nanoparticles that are synthesized by fast, nonequilibrium processes in solution,[25,26] are isomeric (even-proportioned) and rather uniform in size, shape and morphology. Such particles typically grow without developing any large, structure-spanning crystal defects, and therefore our model focusses on local atom dynamics, assuming exact registration of the atoms with the crystal structure. In Section 2, we outline the model details and provide further discussion of the underlying assumptions. In this work, we study breakup on nanowires using this MC dynamical approach. We assume FCC lattice structure, of cubic (non-primitive) cell spacing $a$, typical of noble metals. Surface atoms can hop on the surface, as well as detach, and atoms from the surrounding medium can attach to the particles' surfaces. Motion in the medium is random diffusion, but the hopping and detachment of crystal-structure atoms are carried out according to probabilities set by Boltzmann factors corresponding to free-energies that are set by the atom's neighborhood, as specified in Section 2.

Section 3 presents the results of our modeling and their discussion. An illustration of the type of results that our model yields, is offered in Figure 1. Here, snapshots at different MC times of the breakup process are shown for two different choices of the initial nanowire "diameter" (details in Sections 2 and 3), and one can see examples of that long-lived non-isomeric ("dumbbell") nanoparticles can be formed for certain parameter choices, and that nanoparticle ends can drive the breakup process to form more isomeric nanoparticles due to their tendency to round up faster. All the fragments ultimately round up due to the process dynamics, for large enough times (not shown in the figure). Figure 1 shows an example for a nanowire that reached steady state with the surrounding "gas" of atoms. Our model also allows modeling of situations — not considered in this work — in which the degree of availability of atoms in the surrounding medium can drive growth or dissolution of the initial nanowires, as processes ongoing in parallel to morphology changes due to breakup/instability. Notably, there are experiments[9] with semiconductor nanowires that are highly crystalline and also are in the growth-inducing environment during their breakup process. We expect, based on earlier studies for other processes, that most of our conclusion are relatively general. We note that molecular-dynamics studies of the "nano-thread" breakup process for temperature somewhat above nanowire melting have also been reported,[11] as were results in the premelting regime.[12] Evidence was found of that, in the latter regime development of large defects (stacking faults in the core of the nanowire) can drive the breakup. Our model, however, applies when the structure is crystalline, and our approach is mesoscopic because we use probabilistic model description, which cannot reproduce truly atom-scale effects, but has the advantage of allowing to treat *larger system sizes* not reachable in atomistic simulations. Section 4 offers a summary of our results.

## 2. KINETIC MONTE CARLO MODEL

The utilized kinetic MC approach was developed[18,19] to model shape selection in solution synthesis of isomeric nanocrystals for catalysis and other applications.[25] It was then also used to model surface- and nanotube-templated growth[21,24] of nanoparticles and other nanostructures, as well as sintering of the former.[22,23] There are several accounts of the model in the literature, including in this journal, the latter in our work addressing CVD growth[22] and then dissolution of nanoparticles on carbon nanotubes of interest in nanoscale thermometry.[27] In this section we describe the model with emphasis on details relevant to the considered problem of nanowire instability.

Nanostructure growth and dissolution usually involve fast, nonequilibrium processes driven by matter imbalance. Atoms (or molecules) are assumed to constitute a diffuser "gas," and they can attach to or detach from the nanostructure. Those atoms that are part of the nanostructure locally hop to other, vacant crystal sites, detach (to rejoin the "gas"), and can reattach. To make such mesoscopic models numerically manageable for structures large enough to be of relevance for comparing with trends observed experimentally, a key assumption is made that attached atoms are "registered" with the underlying lattice structure. For definiteness, let us consider a monoatomic structure with coordination number $m_c$ and lattice spacing $a$.

The external "gas" atoms diffuse by off-lattice random-angle hopping, with fixed-length steps that are a fraction of $a$ (here we took $a/\sqrt{2}$). They are attached at unoccupied lattice sites that are nearest-neighbor to the nanostructure atoms, provided they hop into a Wigner-Seitz cell centered at such a site. The precise "registration" of the nanostructure atoms with the crystal lattice ensures[18-22] that we are considering morphologies of relevance for an important class of crystalline nanostructures: those synthesized by fast nonequilibrium techniques, which includes isomeric nanocrystals.[25] These nanostructures do not have structure-spanning defects that can drive their shape selection by preferential growth of certain crystalline faces. Thus, the present model has been most successful for describing shapes in synthesis of isomeric nanocrystalline particles, which are even-proportioned. In reality, of course, large defects are dynamically avoided/not nucleated at the microscopic scales. However, the "exact registration" rule phenomenologically imposes the same property in mesoscopic modeling. For a certain range of sizes, nanocrystals then grow with certain model-parameter-dependent proportions, and, while thermal fluctuations are present at their surfaces, their shapes are to a good precision bound[18,19] by lattice planes of symmetries similar to those in the equilibrium Wulff constructions.[28-30] Particle shapes for standard lattice symmetries[18,19] were found, consistent with the experiments for metal/oxide nanocrystals[31] and core-shell noble-metal nanoparticles.[32,33]

However, we note that the present model has also been useful in exploring shapes of nanostructures other than isomeric nanoparticles, including surface-templated nanopillars,[20,21] and, notably, various aspects of bridging and merger of nanoparticles during their sintering.[22,23] Furthermore, by enclosing the system (nanostructure and "gas") in a "container" either with fixed concentration of atoms maintained in a thin layer as its walls or with reflective walls, one can effectively model the surrounding medium that can not only cause the nanostructure to grow (effective supply of atoms) or dissolve (effective removal of atoms), but also reach a steady state with it surrounding. Here we focus on the latter regime, which only involves shape/morphology changes, and the latter are slow as compared to growth or dissolution. Thus, while the model has its limitations, we expect it to provide useful information as described in the Introduction: the extent to which observed nanowire instability can emerge in a situation where no manifest liquid-like effects are involved (no premelting of the surface), and where the breakup is not predetermined by the dynamics of large internal defects. We can focus on dynamics driven only by the net surface transport of matter (surface diffusion) and/or steady-state exchange of matter (atoms) with the surrounding "gas." As mentioned earlier, potentially the model can also be used to study instabilities with simultaneous growth or dissolution, though this is outside the scope of the present work.

Turning to model details, we note that, each crystal-lattice atom, if it is not fully surrounded, can hop to vacant nearest neighbors (more generally, next-nearest-neighbor hopping, etc., can also be considered[18]), with probabilities for specific moves proportional to temperature-dependent Boltzmann factors. A unit MC time step is defined as a random sweep through the system, such that diffuser-gas atoms are moved once *on average*, whereas lattice atoms have one hopping *attempt* on average. Movable lattice atoms have coordination numbers $m_0 = 1, \dots, m_c - 1$, and they actually attempt to hop with probability proportional to $p^{m_0}$. Here the activation free-energy barrier is $m_0 \Delta > 0$, and $p = e^{-\Delta/kT} < 1$. If the move takes place, the atom ends up in one of its $m_c - m_0$ vacant nearest neighbor sites or is put back into it original site, with the probability proportional to the inverse of a free-energy change Boltzmann factor, $e^{m_t|\varepsilon|/kT}$ (normalized over all the $m_c - m_0 + 1$ targets), with $\varepsilon < 0$ measuring the free-energy of the binding at the target sites. Note that target-site coordination in the resulting configuration can be $m_t = 1, \dots, m_c - 1$ for hopping, and $m_t = 0$ for detachment, where in the latter case, newly detached atoms rejoin the free diffuser gas when they are next "probed" if randomly chosen during MC sweeps.

In each unit-step MC sweep a sufficient number of random selections of atoms is made so that *on average* each atom is "visited" once per sweep. The system is then probabilistically evolved as follows. If the atom is in the gas, a random-direction diffusion hop is carried out; if the final position is within a unit cell near a crystal, the atom becomes crystal-lattice registered as part of the nanostructure. If the atom is in the crystal and is not blocked, it is moved with probability determined by an activation free-energy barrier; the end position is selected according to the Boltzmann factors as described earlier. The end position can be either a part of a crystal, i.e., connected to other atom(s), or the move can leave the atom, and/or (some of) its neighbors, not connected to the crystal. Such disconnected atom(s) are all immediately reclassified as located in the gas for their future hopping moves.

In our case, the nanowire was surrounded by a reflective container, and the total number of atoms was therefore unchanged with time. The container surface was at distances from the initial nanowire faces that were typically 10 times larger than its effective radius. After a short transient time, enough atoms were detached to form a "gas" (of diffusers) in steady state with the nanostructure. This was a small fraction of the total number of atoms, and therefore, for convenience the container was assumed initially vacant. For Figure 1, the initial nanowire and the surrounding container were oriented along a (100)-type direction of the cubic lattice structure of FCC ($m_c = 12$). Initially, the nanowire contained (A) ≈ 115700 atoms, of which ≈ 101700 remained in the nanostructures at all the shown later times, after the formation of the gas; and (B) ≈ 100300 atoms, of which ≈ 86000 remained. The initial nanowires length, $L$, in our various simulations was a large multiple of the lattice spacing $a$, and effective nanowire diameter, $d$, was a multiple of $a$, ranging from 10 to 20. In the system presented in Figure 1, we assumed that the cross section of the initial nanowire was octagonal, limited by 4 + 4 (100)- and (110)-type lattice planes, separated by $d$, and then the resulting cylinder was "filled" with the FCC lattice sites, to include as many lattice planes as fitted in, with the lattice matching the appropriate orientations. *Other orientations* with respect to the crystal planes and also different cross-section shapes of the initial nanowire were also considered; see Section 3.

The surface atom mobility is related to the surface diffusion coefficient, which is in turn proportional to $p$, controlled by the aforementioned activation free-energy scale Δ, with

$$p = e^{-\Delta/kT}. \tag{4}$$

The other introduced free-energy scale, $\varepsilon$, reflects local binding, and we define

$$\alpha = |\varepsilon|/kT. \tag{5}$$

Earlier numerical studies[18-24] indicate that within the present modeling framework typical nonequilibrium nanostructure morphologies are maintained for a range of mesoscopic sizes if we use reference values $\alpha_0$ and $p_0$ comparable to 1. Temperature, $T$, can then be increased or decreased by varying $\alpha$, which is inversely proportional to it, provided $p$ is adjusted according to

$$p = (p_0)^{\alpha/\alpha_0}. \qquad (6)$$

Our prior experience with FCC nanocrystals and other structures has suggested the use of the following reference "intermediate temperature" values for our particular model MC rules: $\alpha_0 = 1$, $p_0 = 0.7$, to relate $p$ to the choice of $\alpha$. In this work we actually focused on results for the following two cases: elevated-temperature "hot" systems, $\alpha = 0.9$, e.g., Figure 1, and reduced-temperature, $\alpha = 1.2$, "cold" systems (see Section 3).

## 3. REPRESENTATIVE RESULTS AND FINDINGS

### 3.1. *Stages of the Breakup Process*

In this section we describe and illustrate various stages and features of the nanowire breakup process as observed in our numerical modeling. The present subsection summarizes various process stages. The initial, fast establishment of the steady state with the "gas" of atoms in the reflecting-wall container (the gas atoms in our case originate by evaporation from the nanostructure) is accompanied by the rounding of the initial nanowire shape. Specifically, various crystal faces remain well-defined, but noisy (not fully smooth) due to fluctuations, which is reminiscent of the same effect in synthesis of nanoparticles.[17-24] However, the proportions of the cross-section are significantly modified: It de facto become "isomeric," i.e., the nanowire shape becomes even-proportioned, but only in the cross-section. Therefore, the precise form of the initial cross-section is not that important. This feature is illustrated in Section 3.3.

Later stages of the process are mostly sensitive to the total number of atoms per long segment of the nanowire, and, of course, to the temperature (and various choices of the free-energies and other model parameters), as well as to orientation of the crystalline structure with respect to the initial nanowire shape. The process proceeds via two mechanisms. Necks form along the nanowire, and at the same time its ends can drive break-off of isomeric nanoparticles. Necks usually narrow down and result in breakup that leads to isomeric nanoparticles. However, for some parameter choices there is a small, but non-negligible likelihood that some of the necks will be long-lived, and then dumbbell-like shapes can break off the nanowire. These shapes can remain stable for extended times, even though eventually they also round up (but without further dividing into smaller parts). We note that long-lived multi-bulb "stable" structures are even less probable as compared to two-blub dumbbells. Here by "probabilities" of the long-surviving dumbbell and more complex non-isomeric structures we mean the fraction of the MC "histories" for which they are found.

### 3.2. *General Properties of the Breakup Process Along the Nanowire*

As described before, after the fast establishment of the steady state with the "gas" in the container, the nanowire develops a wave-like neck structure. This is a random-fluctuation and local-dynamics driven process, but certain analogies can be found with liquids. Figure 2 illustrates a process similar to that in Figure 1, but for a different crystalline orientation, (111)-type, with respect to the nanowire shape. However, in Figure 2 we picked an instance of time, image (a), at which one can see an obvious more or less uniform wave-like "bulging," except near the ends of the nanowire, where the dynamics is different. At later times, the configuration in Figure 2 breaks up into isomeric and dumbbell fragments. Interestingly, both dumbbells seen in image (d) are long-lived and they eventually only round up, rather than break up.

The Inset in Figure 2 shows a typical shape of a fragment that ultimately ends up as an isomeric nanoparticle (without going via the dumbbell stage) formed between by two necks. This was obtained for a longer nanowire, for which the end-region dynamics was artificially "frozen" (five layers at each end were not evolved under the MC dynamics, but kept in the initial configuration) to minimize end-effects. The center-to-center distance of such fragments fluctuates, but on average it is $\Lambda \approx 7.1 r_0$, provided we identify $r_0 \simeq d/2$. However, possible similarity with the breakup of liquid jets should be taken with caution, cf. Equations 1, 2, 3. This is further enforced by that, we only got four end-stage nanoparticles, because both dumbbells shown in image (d) were found not to break up for later times of the MC run, whereas the short-time wave-like instability, marked by arrows in image (a) would suggest seven fragments.

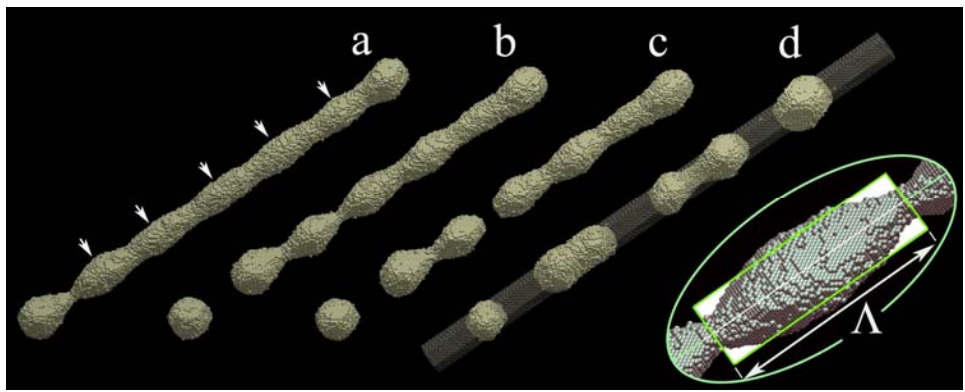

**Figure 2.** Random MC run showing snapshots at several times, for a nanowire with its axis along the (111)-type orientation of FCC, with model parameters $\alpha = 0.9$, $p = 0.725$, $L = 240a$, $d = 12a$, with configurations labelled (a), (b), (c), (d), showing the system state after $t = 2.05, 2.90, 3.05, 4.60 \times 10^6$ MC time steps, respectively. The image labelled (d) also outlines the initial nanowire, which was bound by six (110)-type FCC faces (the cross-section here is hexagonal). The arrows highlight the onset of the approximate periodicity of the nanowire distortion in the configuration shown in image (a). The Inset shows a typical segment between two necks for a longer nanowire, $L = 400a$, with "frozen ends" (see text). The white rectangle highlights a region that will ultimately form a separate nanoparticle.

The dynamics of breakup process is random not only on the few-atom scale of local dynamics that proceeds by Boltzmann factor-controlled MC moves, but also globally. This is illustrated in Figure 3, where we show the same system as in Figure 2, but for different random number generator initializations. We note that, while the initial number of atoms here was approximately 113300, after atoms were lost to the "gas," for all the runs, including the ones shown in Figure 2, 3 and several not shown, the resulting nanostructures during the steady-state shape-change stages of the dynamics retained approximately 86300 atoms. While the amount of matter lost into the "gas" upon equilibration was the same for all the runs, the random nature of the final configurations, of the timings of the various breakup, and of other restructuring stages is clearly seen. Specifically, the number and nature of the fragments (nanoparticle, dumbbell) is different, and even when their count is the same, the breakup scenarios and the resulting fragment sizes differ. The fragment in the Inset in Figure 3 assumes a well-defined dumbbell configuration at $t = 3.95 \times 10^6$, and then remains practically unchanged (up to fluctuations within 1–2 atomic layers at the surface) up to the $t = 5.25 \times 10^6$, for which is it shown in the Inset. By that time, all the other fragments for all the depicted runs fully "rounded up" to isomeric nanoparticles (not shown) and the MC simulation was stopped. We note that occasional emergence of dumbbells in nanowire breakup, seen in our MC modeling, is consistent with the experimental observations[1,6,8,9] and continuum model,[13] where the latter reported that nanowires can "neck down to the next stable radius."

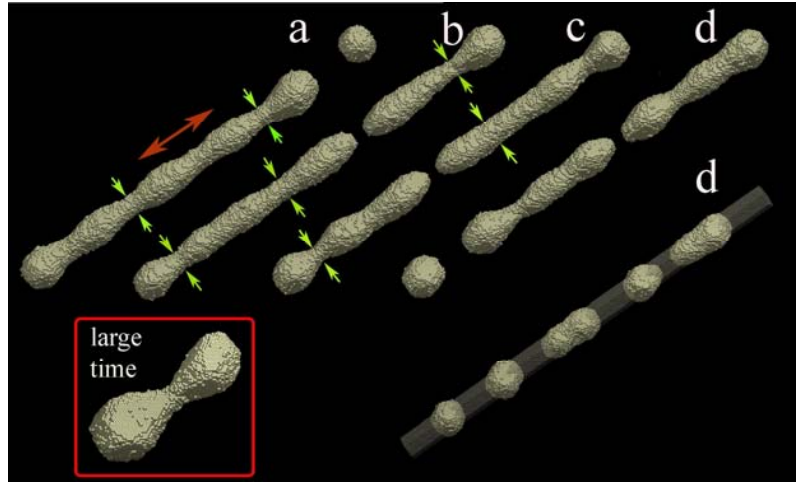

**Figure 3.** Four additional, different random MC runs, labelled (a), (b), (c), (d), for the same system as in Figure 2. Images (a), (b), (c), (d) show snapshots for different times for each run, respectively, $t = 3.05, 1.84, 2.42, 2.25 \times 10^6$. These are the times when the initial necking was crossing-over into the first breakup(s) along the nanostructure, and these times randomly fluctuate between the realizations. The arrows for images (a), (b), (c) show the locations of the later breakups, whereas the lower image (d) shows a later-stage, $t = 3.05 \times 10^6$, configuration obtained as the earlier configuration (upper image) was further aged (superimposed on the outline of the initial nanowire). For runs (a), (b), (c), these

- 13 -

later-stage, arrow-marked breakups are *all* completed by the respective times 3.56, 3.49, 3.39 × 10$^6$. These, as well as intermediate breakup times are therefore also relatively random. The Inset is a large-time image of the neck marked by the double-arrow in image (a). It remains rather stable after all the other breakups are already fully competed.

Various realizations of the type shown in Figures 2 and 3 yield isomeric nanoparticle "droplets" of differing sizes. For those that were well-formed, we got the average size estimate of $\langle R \rangle \approx 10.5a$, with the r.m.s. deviation $\Delta R \approx 0.16 \langle R \rangle$, where we use the notation introduced in the discussion following Equation 3. Therefore, our estimate of the fluctuations in typical particle sizes, ~ 16%, is comparable to experimental values for noble-metal nanowire breakup.[1,4] The average distance between centers of nearby well-formed nanoparticles was estimated as $\langle \Lambda \rangle \approx 54a$. On the other hand, the fraction of atoms remaining after the "gas" was established, see the discussion of Figure 3, can be used to estimate the effective average "jet" radius as $\langle r_0 \rangle \approx 5.2a$. Thus, here the ratio $\Lambda/r_0$, averaging ~ 10.4, is somewhat larger than continuum-model and liquid-jet results, consistent with experiment.[4]

### 3.3. *Nanowire Geometry Effects: Initial Shape, Crystalline Orientation*

In this subsection we address certain "geometrical" effects related to that, the nanowire can have initially different cross-sections, and also have its crystalline structure differently aligned with respect to its initial shape. Let us first discuss the effects of the initial cross-section shape. We already commented earlier that, the tendency to have "isomeric" surface features seems to apply not only to whole nanoparticles but also to the initial distortion of a long nanowire's cross-section — rounding as a tendency to become primarily bound by certain crystalline faces, in our case of FCC, by (100)-, (110)- and (111)-type. Therefore, in practice the actual preparation of nanowires of the type that can be described by our model, i.e., those the shape of which is not dominated by the effects of large internal defects, should already result in such shapes. However, in simulations we can in principle consider arbitrary initial shapes. An illustrative example of the initially slab-shaped nanowire is presented in Figure 4. We see that the MC dynamics causes rapid onset not only of local surface fluctuations, but also of larger-scale transport of matter that results in rapid restructuring whereby the cross-section becomes "isomeric," approximately octagonal, bound by noisy versions of the (100) and (110)-type faces. The latter, rounded nanowire then undergoes the breakup into ultimately isomeric nanoparticles. Note that here, the surrounding container, selected to match the initial nanowire, retains the memory of that slab-like shape, but diffusional mixing of the "gas" of atoms suffices to erase the effect of the container cross-section on the later nanowire breakup stages.

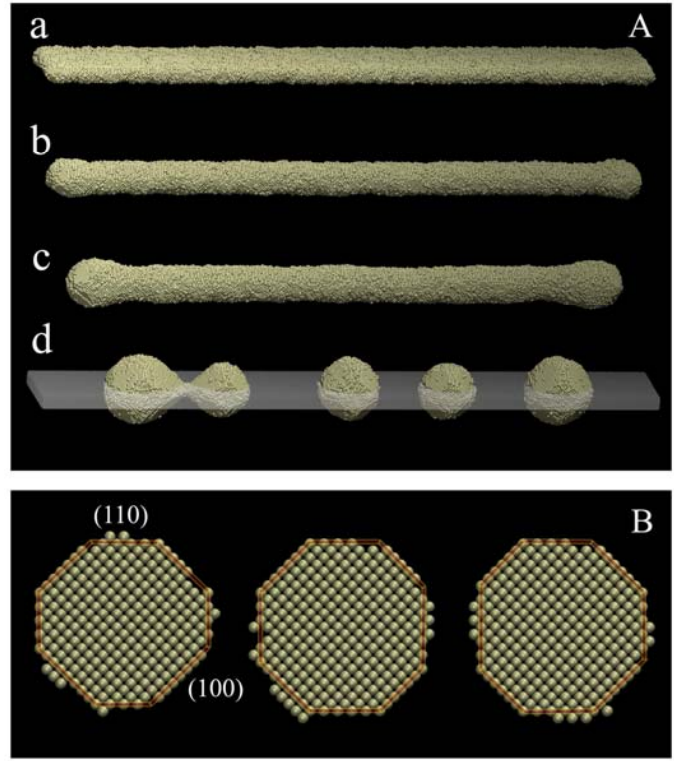

**Figure 4.** (A) Random MC run showing snapshots at several times, for an initially slab-shaped nanowire with its axis along the same, (100)-type orientation of FCC as in Figure 1, with configurations labelled (a), (b), (c), (d) showing the system state after $t = 0.01, 0.20, 1.50, 7.90 \times 10^6$ MC time steps, respectively. The system parameters were the same as in Figure 1, except for the initial shape, outlined in image (d), which was a slab of dimensions $4a \times 30a \times 240a$, i.e., of the same length and approximately the same volume as the configuration in Figure 1. (B) Three cross-sections of the "isomerically rounded" nanowire, shown

in panel (A), image (b), at 1/4, 1/2, and 3/4 of its length (from the left edge), depicted left to right, respectively. The octagonal contours were added to guide the eye.

In principle, we can consider an arbitrary-orientation long-cylinder shape of a fixed (not necessarily round or hexagonal or octagonal, etc.) cross-section and oriented along an arbitrary direction with respect to crystal-symmetry axes. The considered MC dynamics is expected to cause "isomeric rounding" in any case. However, if the selected orientation is not consistent with "rounding" to a cylindrical shape bound by the favorable surface planes, then the product will not be straight. Rather, we could expect that stepped or potentially even helical shapes will emerge (specifics depend on the material and its crystalline structure), made primarily of noisy versions of crystalline faces. Indeed, our preliminary studies for FCC suggest that fragments of higher-index faces will also emerge. We note that, our model excludes large internal defects and their dynamics. It is possible that the initially "mismatched" nanowires will actually acquire the preferred surface structure by developing large defects. The present model therefore can lead to unphysical results in such cases. We note, however, that the structure of various non-linear (stepped, helical, etc.) nanowires that have been experimentally realized,[5,34,35] have been found or theoretically argued to be either highly crystalilne[5,34] or large-defect dominated.[12,35] Therefore, we plan to explore such structure emergence of the former in the framework of the considered model in our future work.

### 3.4. *Physical Processes and Morphological Effects that Control Nanowire Breakup*

Perhaps the most interesting effect, confirming experimental findings[9] for single-crystal Au nanowires, of the crystalline structure orientation is observed when the initial nanowire long axis is along (110) of the cubic (non-primitive) FCC lattice structure. The is illustrated in Figure 5. Here we took a nanowire with length somewhat shorter than that encountered earlier, to be able to clearly show its dynamics. The initial nanowire contained 57000 atoms. Low-temperature dynamics, Figure 1(A), results in atoms lost to the "gas," to leave approximately 55000 atoms in the shown cluster. Higher-temperature dynamics leaves less atoms in the cluster, ~ 38000. In both cases the nanowire undergoes the expected "isomeric rounding," and the shape of the ends is illustrated in the Inset. However, there is no breakup into nanoparticles. Instead, the whole structure gradually shortens to ultimately result in a single isomeric nanoparticle. For the shown time, the lower-temperature nanowire already shrank to 86% of the selected initial length, $L$, whereas the higher-temperature one shrank to 47% of $L$. We also checked numerically that, this length, $L$, would suffice to observe breakup into 3 to 4 fragments, for nanowires of cross-dimensions $11a - 14a$ for the selected time, were they initially along (100) or (111). Experimental work[9] offers additional discussion in terms of surface binding energies, of their finding that nanowires oriented along (110) are more difficult (require longer times) to break up than those with other orientations.

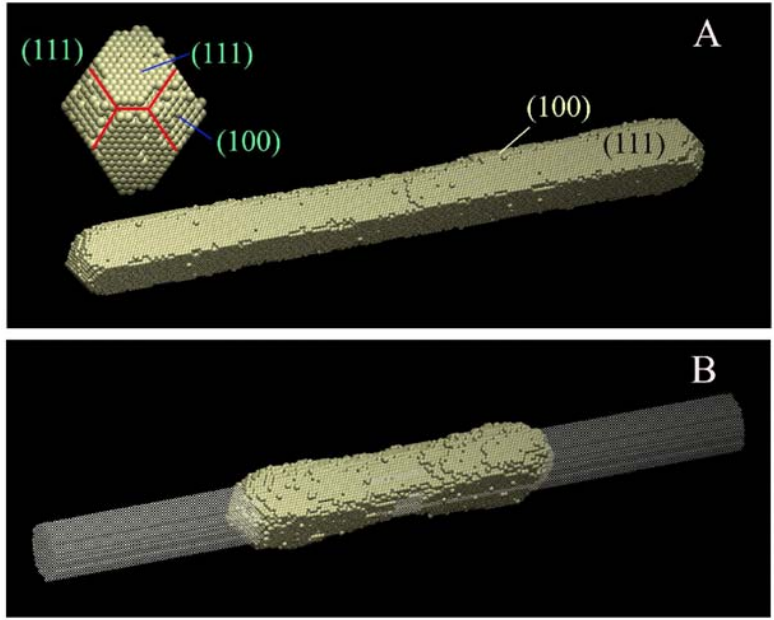

**Figure 5.** (A) Random MC run showing a snapshot at time $t = 6.0 \times 10^6$ of a nanowire oriented along the (110) direction, of initial length $L = 150a$, and cross-dimension of size $d = 11a$. This simulation was carried out for a reduced temperature, $\alpha = 1.2$, $p = 0.652$. The Inset shows the frontal view of the left end, with red lines added to highlight lattice edges. However, there is no breakup process. (B) The same for a higher temperature, $\alpha = 0.9$, $p = 0.725$, which was assumed earlier for simulations illustrated in Figures 1–4. The initial nanowire (for both temperatures) is outlined in (B).

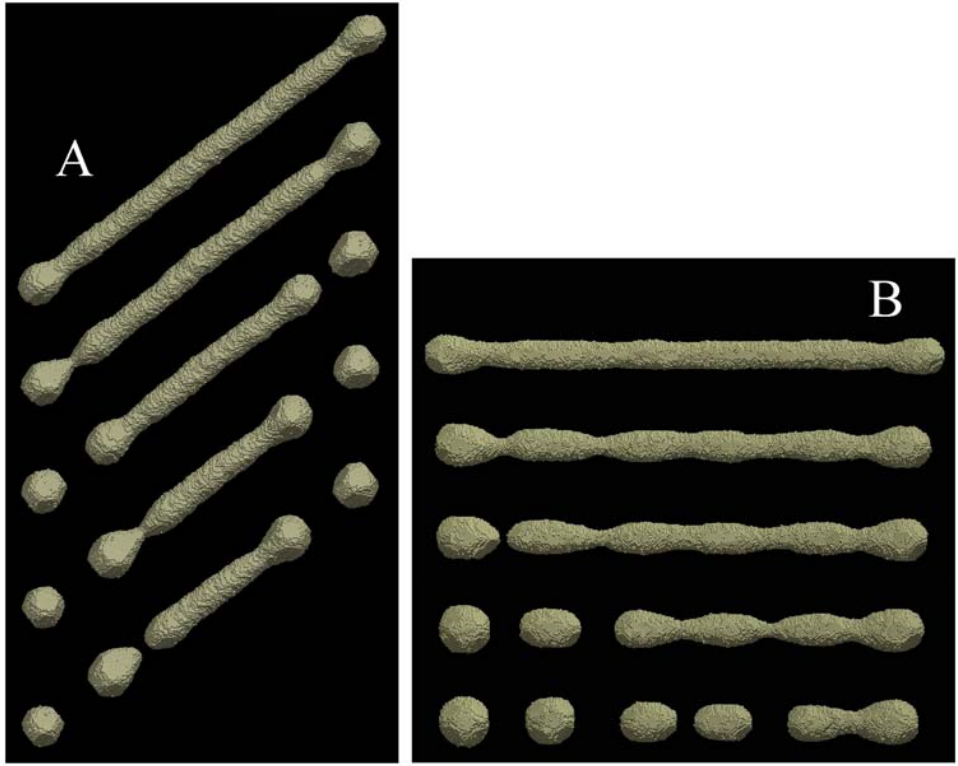

**Figure 6.** (A) Random MC run showing a snapshot of the same system as in Figure 2, but for a lower temperature, $\alpha = 1.2$, $p = 0.652$, Note that the nanowire here is longer than in Figure 5(A), and its end-driven breakup is shown for times $t = 5.0,\ 10.0,\ 15.0,\ 20.0,\ 21.15 \times 10^6$, top to bottom. (B) The same, lower-temperature simulation for the system shown in Figure 1(B), for times $t = 4.0,\ 9.0,\ 10.0,\ 12.0,\ 14.0 \times 10^6$.

In earlier studies of nanoparticle sintering,[22,23] it has been reported[23] that necks between nanoparticles can coarsen or dissolve, or even in some cases remain stable for extended times, depending on the local neck geometry and, importantly, on the competition of the net surface diffusion with transport through and exchange of matter with the gas of atoms. This competition of different transport mechanisms depends on the temperature, but also on the specific local geometry of the crystalline faces that are present in the neck regions. We note that, while surface diffusion is effective in smoothing out small fluctuations, exchange of matter with the gas is the process most sensitive to surface curvature and, similarly to the the "Laplace-pressure" considerations for liquids, it is crucial for the emergence of the necks. It transpires that for the (110)-oriented nanowires necking is not present for both considered temperatures, the higher of which was the same as for (100) and (111) nanowires, cf. Figures 1–4. However, as we lower the temperature, here by taking $\alpha = 1.2$, it turns out that the (111) nanowire also losses the ability to develop necks within the nanowire, whereas internal necking is still observed for the (100) case. This is illustrated in Figure 6, which also shows that end-effects, discussed in the next subsection, can still drive nanoparticle formation, here for (111).

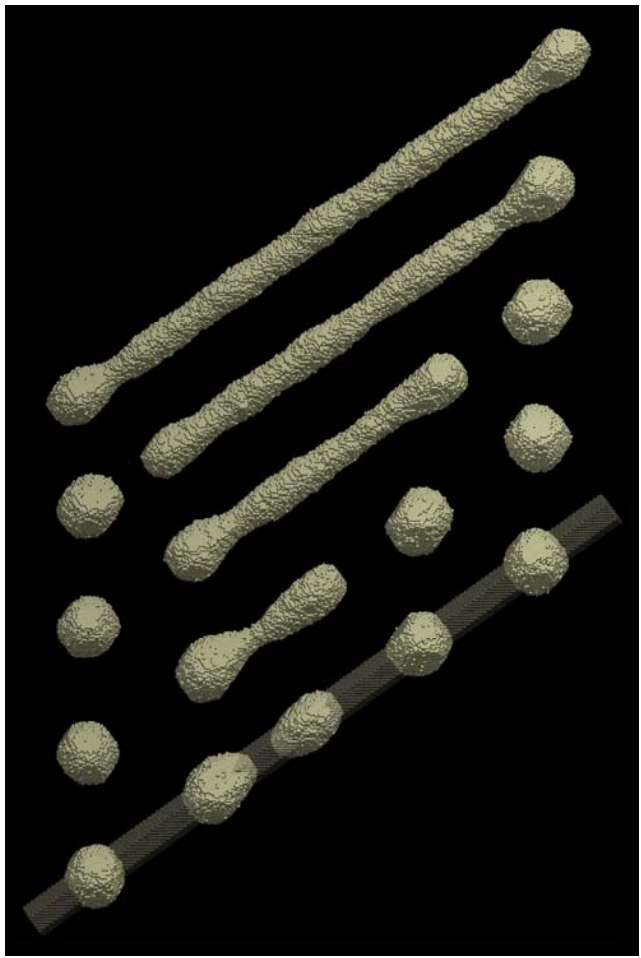

**Figure 7.** Random MC run showing a snapshot of the same system as in Figure 2, keeping the higher temperature, $\alpha = 0.9$, $p = 0.725$, but with a smaller initial

nanowire cross-section size, $d = 10a$, as explained in the text. Here the atom detachment dynamical steps were excluded (see text). The resulting nanostructures are shown for times $t = 2.0, 4.0, 6.0, 8.0, 8.4 \times 10^6$, top to bottom, and the initial nanowire is outlined in the bottom image.

One of the advantages of the present modeling approach is that, certain dynamical processes can be switched on and off to explore interesting physical effects. We already mentioned the "freezing" of the dynamics of the nanowire ends, mentioned in connection with the Inset in Figure 2. Les us now consider another modification of the dynamics whereby we change the rules for attached particle motion to exclude detachment. Without describing in detail the modified dynamical rules, we only present, in Figure 7, an illustrative result that enforces the conjectures made above regarding the roles of surface diffusion vs. exchange of matter with the gas. Here we took the same, higher-temperature system as in Figures 2 and 3. However, instead of lowering the temperature (Figure 6A), we eliminated dynamical moves of particle detachment. We also took a smaller-diameter initial state, in order to have approximately the same number of atoms in the crystalline nanostructures as before (because with detachment allowed, a fraction of atoms was lost to the gas). The key observation is that, here, like in the case of the lower temperature (Figure 6A), the instability along the wire is suppressed by switching off atom detachment, while keeping surface diffusion. However, the end-effects, discussed in the next subsection, are still effective at breaking nanoclusters off the nanowire for long enough times.

### 3.5. *End-Effects in the Dynamics of Nanowire Breakup*

Inspection of Figures 1–3, 6, 7 provides an illustration of that, nanowire ends can drive breakup by generating a sequence of nanoparticles that are more likely to be isomeric than fragments originating from the interior region. Here, we will describe the dynamics of this process. Figure 8 shows results for a short nanowire with a frozen end, which illustrates that the process of the formation of nanoparticles breaking off the other end is random (takes different times), generally shortens the remaining nanostructure, and the nanoparticles that broke off have a certain distribution of sizes. To further elucidate this process, let us consider an example shown in Figure 9, which depicts the top portion of another, lower-temperature nanowire. As the nanowire undergoes initial "isomeric rounding," its ends, which have more crystalline faces present and can also recede into the main structure, tend to form a more bulbous portion than the interior region. The distortion in the nanowire shape then includes a neck. However, generally necks need not always narrow down. Their dynamics is actually controlled by both the net surface diffusion and exchange of atoms with the gas, as was noted in earlier studied of sintering.[23] Apparently, the asymmetry of the *end-induced* structure makes the diffusional-transport effects more profound than in the case of, for instance, internal dumbbells, etc. Results with atom detachment excluded, exemplified in Figure 7, illustrate this interesting observation.

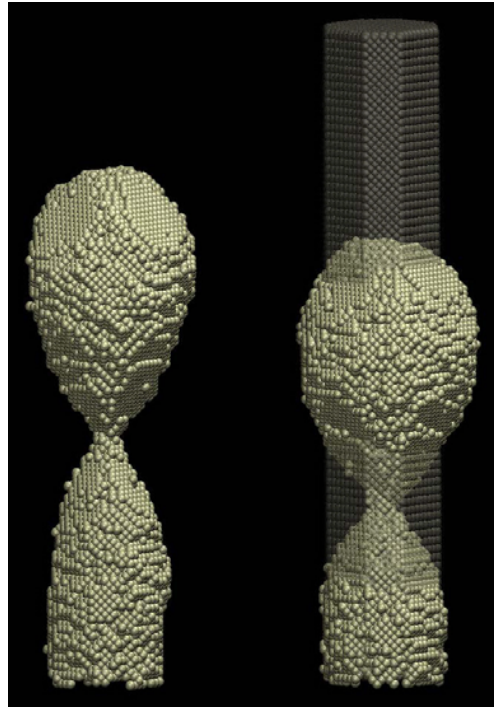

**Figure 8.** Two different random MC runs showing snapshots (depicted vertically) of the system similar to that in Figure 1(A), at the same elevated temperature, $\alpha = 0.9$, $p = 0.725$, with initial nanowire cross-section size, $d = 12a$. However, here the initial nanowire (outlined on the right) length was much shorter, $L = 75a$, and furthermore the crystal atom dynamics in the 5 lowest planes was frozen (though gas atoms still attach at the bottom face, which makes it look noisy). Both images show the nanowire at the instance of time right before

a nanoparticle breaks off it, $t = 2.43 \times 10^6$, and $t = 4.11 \times 10^6$, for the left and right runs, respectively.

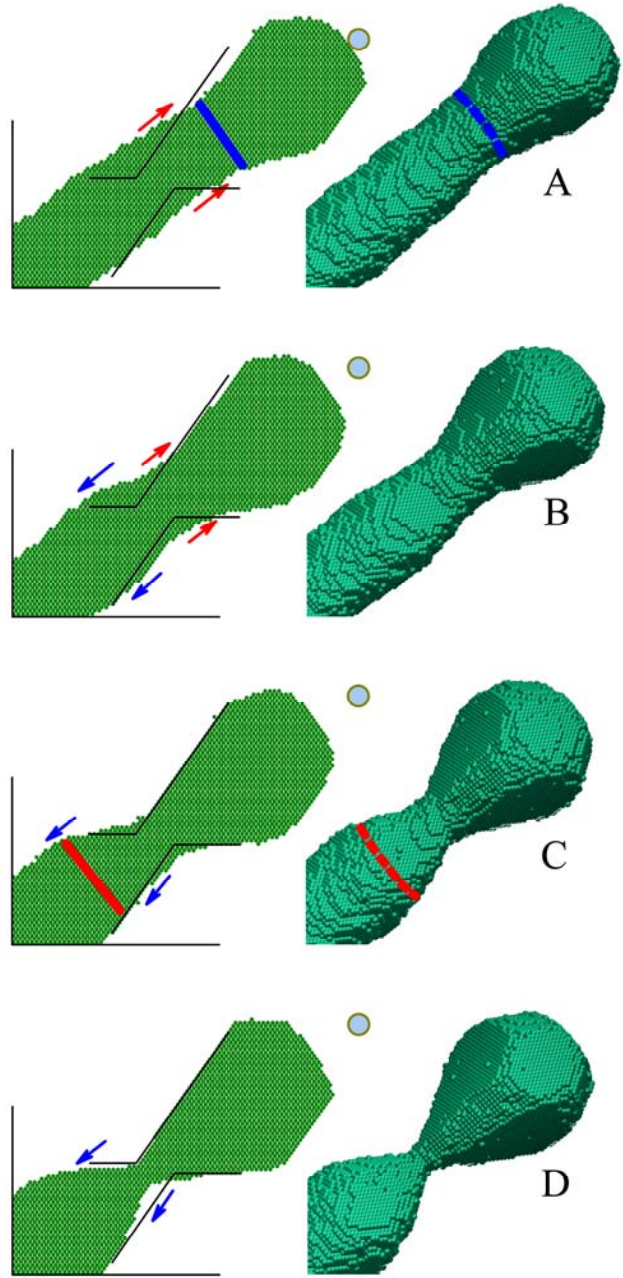

**Figure 9.** Random MC run showing snapshots (right images) and projections onto a (110)-type plane (left images) of the top portion of a system similar to that in Figure 6(A), at the same reduced temperature, $\alpha = 1.2$, $p = 0.652$, with the initial nanowire cross-section size $d = 14a$. The initial nanowire length was short, $L = 100a$, and furthermore the crystal atom dynamics in the 5 lowest planes (not shown) was frozen. The large light-blue, olive-circled dot is at a fixed position in all the panels. It was added to provide a point of reference indicating how the nanostructure as a whole shortens during the process. Panels (A), (B), (C), (D) show the top of this nanowire at times $t = 7.0, 12.0, 20.0, 25.0 \times 10^6$, respectively, whereas the nanoparticle fully breaks off at $t = 25.18 \times 10^6$, and leaves behind nanowire of lengths $52a$ (between the rounded top and flat bottom). The blue lines in (A) mark the onset of the neck. The red arrows in (A) and (B) indicate the direction of the initial diffusional flux of matter. The blue arrows in (B), (C), (D) mark the opposite flux that sets in at later times. The net effect of this, as well as of exchange of atoms with the gas is not only tightening of the neck, highlighted by the two broken black lines (which are in the same locations in all the panels), but also recession of the neck into the nanowire, driven by the emergence of a curved structure past the neck, marked by the red lines in (C).

For end-induced necks, asymmetric transport initially pulls matter into the bulbous end-structure, causing neck-narrowing, and at the same time that structure recedes into the nanowire (the nanowire shortens). However, as the neck narrows, the opposite pull later sets in on the other side of the neck, into the interior of the nanowire. This is illustrated in Figure 9. The overall effect is that typically, the whole nanowire continues to shorten, the neck narrows down, and ultimately a nearly isomer nanoparticle breaks off at the end of the nanowire. However, we emphasize that the process is random, and there are significant statistical fluctuations in the resulting configurations. Indeed, our runs depicted in Figure 1(B), image (d); Figure 3, image (d), Figure 4(A), image (d); and Figure 6(B), the bottom image, demonstrate that the end dynamics can also result in the breakoff of dumbbell structures, i.e., the neck(s) formation in the interior can occur faster than the end-induced necking.

## 4. SUMMARY

The utilized MC model is mesoscopic and therefore is presently limited in that it can provide only approximate description of specific generic behaviors. Here it reproduces and elucidates several features of the breakup process, as detailed specifically in Sections 3.2-3.4, which are consistent with experimental observations. For more quantitative modeling one would either have to connect model parameters and MC time scales to results of microscopic studies, or to experimental data. The former, truly microscopic modeling, is presently not available without additional phenomenological assumptions, within the DFT, molecular dynamics, or other approaches, in a formulation that can reproduce crystallization starting from the formation of amorphous few-atom embryos to their growth into crystals. The latter, detailed time-dependent experimental data, are presently very limited, though some results have recently become available for the time-dependent dynamics of particle dissolution.[24,27]

Our model actually primarily addresses those experimentally observed features that cannot be described by continuum mechanisms. Thus, we study the regime for which analogy with liquid-jet breakup is not directly applicable and most similarities are likely coincidental. Specifically, for our case, without actual surface "premelting," surface diffusion alone was found not to be particularly effective at inducing the breakup along the nanowire, except via end-effects. When surface diffusion is supplemented with exchange of matter with the surrounding medium, the nanowire breakup sets in internally and also via the end effects, though the internal breakup efficiency is strongly dependent on the initial nanowire morphology, temperatures, etc., whereas end effects are uniformly more effective at breaking off isomeric nanoparticles. Some of our estimates of fragment and other characteristic dimensions in nanowire breakup are consistent with experimental observations,[1,4] and, importantly, differ from values for liquid-jet breakup. We also confirm the experimental finding[9] that (110) nanowires are more difficult to break up than those long (100) or (111)-type directions.

Initially, the nanowire cross-section undergoes fast "isomeric rounding". Then, for short times, necks, multi-bulb and dumbbell structures will be present, for parameter values for which the interior breakup is efficient. While the process is largely random, for lager time most necks shrink, and eventually the nanowire structure falls apart into nanoparticles. However, for a small fraction of MC-dynamics "histories" some necks will not shrink but will result in dumbbell and other non-isomeric long-duration structures, though ultimately they will either round up to larger than average isomeric-nanoparticle fragments, or break apart into smaller fragments. All these observations are based on our modeling of many realization of the considered dynamics, for various system lengths, cross-dimensions (with $d$ between $10a$ and $20a$), only a small fraction of which are shown as snapshots in the figures in this work. These results sufficed to reach qualitative, though not very precise estimates of the relative probabilities of those MC "histories" that had configurations with long-surviving non-isomeric structures. Dumbbells, etc., that remain approximately stable for long times (as compared to the times of other ongoing restructuring processes in the system) are of interest because they were experimentally observed, and furthermore, for applications such non-isomeric structures can be stabilized by various means (by modifying the environment).